\documentclass[12pt]{article}
\title{
Fine structure splittings of excited $P$ and $D$ states in charmonium
}
 \author{A.M.Badalian, V.L.Morgunov,\\ Institute
of Theoretical and Experimental Physics,\\
 117218, Moscow, B.Cheremushkinskaya
25, Russia,\\
and\\
 B.L.G.Bakker\\
  Free University, Amsterdam,  The Netherlands}
 \newcommand{\be}{\begin{equation}}
\newcommand{\ee}{\end{equation}}
\def\la{\mathrel{\mathpalette\fun <}}

\def\fun#1#2{\lower3.6pt\vbox{\baselineskip0pt\lineskip.9pt
\ialign{$\mathsurround=0pt#1\hfil
##\hfil$\crcr#2\crcr\sim\crcr}}}
\date{}

\begin{document}
\maketitle

\begin{abstract}

It is shown that the fine structure splittings of the $2\,^3P_J$ and $3\,^3P_J$
excited states in charmonium are as large as those of the $1^3P_J$ state
if the same $\alpha_s(\mu)\approx 0.36$ is used. The predicted mass
$M(2\,^3P_0)=3.84$ GeV appears to be 120 MeV lower that the
center of gravity of the $2\,^3P_J$ multiplet and lies below the $D\bar D^*$
threshold. Our value of $M(2\,^3P_0)$ is approximately 80 MeV lower than
that from the paper by Godfrey and Isgur \cite{1} while the differences in 
the other masses are  $\la 20$ MeV.
Relativistic kinematics plays an important role in our analysis.
\end{abstract}

\section{Introduction}

At present only the $1^3D_1$ and $2\,^3D_1$ states lying above the $D\bar D$
threshold have been identified with the experimentally observed
$c\bar c$ mesons, $\psi (3770)$ and  $\psi (4160)$. Still a large
number of other excited $P$- and $D$-wave states above the flavor
threshold were predicted. Their masses and fine structure splittings
were calculated by Godfrey and Isgur (GI)  already in 1985, in the
framework of a relativistic approach \cite{1}. The properties of the $P$
and $D$ levels in charmonium and bottomonium were also intensively studied
in the nonrelativistic approximation \cite{2,3}.
There is a point of view
that one or more  charmonium $2P_J$ states can be narrow enough
to have a substantial branching ratio to the $\gamma+
\psi(2S)$  channel \cite{4,5} and could play a role in the hadronic
production of $\psi (2S)$ mesons. In particular, the $2P$ states can be
related to the enhancement in the $J/\psi \pi^+\pi^-$ system near
$M=3.84$ GeV observed in \cite{6} (but not confirmed by another group
\cite{7}). Therefore the precise knowledge of their masses is
especially important.

An accurate description of the charmonium spectrum and the fine structure
splittings of the $1P$ level was presented in our previous papers
\cite{8,9}, where, as well as in \cite{1}, relativistic kinematics was
taken into account with the help of the spinless Salpeter equation.
As shown in \cite{9}  the  relativistic corrections to the
matrix elements, like $<r^{-3}>$ determining spin structure, are
large enough, of the order of $\approx 40$\%, therefore the nonrelativistic
approach can not be considered to be appropriate when the spin
structure is investigated.

We shall show here that in the relativistic approach the spin--orbit and
tensor splittings are rather large for the  $P$ states, for
the ground state as well as the excited  levels. This result depends
weakly on the choice of the strong coupling constant
$\alpha_s(\mu)$.  Here $\alpha_s(\mu)\approx 0.36\;\; (\mu=0.92$ GeV)
will be used for all states, but the splittings  do practically not
change if $\alpha_s(\mu)=0.30\;\;(\mu=m=1.48$ GeV) is taken.

The fine structure splittings predicted here appear to be  larger
than those  in GI's paper \cite{1}, especially for the $2P$ and $3P$  states.
The reasons for that will be discussed in Section 3. As
a result the $2\,^3P_0$ mass, $M(2\,^3P_0)=3.84$ GeV in  our case
appears to be approximately 80 MeV smaller than in \cite{1} and this level
lies below the $D\bar D^*$ threshold. The  $2\,^3P_{1,2}$ levels as well as 
the $n\,^3D_J$ states  have mass values close to the GI predictions.
For the first time we also predict large fine structure splittings
for the $3P$ states.

\section{Spin--averaged spectrum}

The relativistic effects in  charmonium are  expected not to be small,
especially for the wave functions and matrix elements which we are mostly
interested in here.  Therefore  to find the spin-averaged spectrum the spinless
Salpeter equation will be solved as already done in several papers
\cite{1,10,11}
\begin{equation}
 \left(2\sqrt{p^2+m^2}+V_0(r) \right) \psi_{nl}(r) = M_{nl}  \psi_{nl}(r) .
 \label{eq01}
\end{equation}
The static interaction $V_0(r)$ will be taken in the form of the Cornell
potential,
\begin{equation}
V_0(r)=-\frac{4}{3}\frac{\tilde \alpha}{r} + \sigma r +C_0
 \label{eq02}
\end{equation}
and the  values of
$\tilde \alpha\equiv \alpha_V(\mu)$, the string tension $\sigma$ and the
pole mass of the $c$ quark will be taken as in our previous paper \cite{8}
while fitting the fine structure of the $1P$    charmonium state,
\begin{equation}
m=1.48 \, {\rm GeV}, \quad \sigma = 0.18 \, {\rm GeV}^2, \quad \tilde \alpha =0.42 .
 \label{eq03}
\end{equation}
The constant $C_0$ in Eq.~(2) is determined from a fit to the spin--averaged 
mass of the $1S$ states, $\overline M(IS)=3067.6$ MeV
 \cite{12}, and from there $C_0=-140.2$ MeV.
In our approach the strong coupling constant $\tilde \alpha$ is not an
independent parameter. It can be connected to $\alpha_s(\mu)$ (in the
$\overline{MS}$  renormalization scheme) through the  relation \cite{13}
\begin{equation}
\tilde  \alpha (q^2) =\alpha_V(q^2)=\alpha_s(\mu)
 \left(1+\frac{1.75}{\pi}\alpha_s (\mu) \right) \; (n_f=3) .
 \label{eq04}
\end{equation}
For our choice of $\tilde \alpha$ and $\alpha_s(\mu)$ (see below)  this
relation will be valid with an accuracy better  than $5\%$.

The chosen parameters (3) can be compared
to the GI  parameters in \cite{1} where $m=1.628$ GeV
 while $\sigma =0.18 $ GeV$^2$ coincides with $\sigma$
in (3).
In \cite{1} a running coupling constant was used for $\tilde \alpha$ 
with the critical value  $\value $\alpha_{cr}= \tilde \alpha
(r=0)=0.60$ which is larger than the constant $\tilde \alpha =0.42$
in our case.  Also in \cite{1} $C_0=-253$ MeV  whereas
 $C_0=-140 $ MeV in  our calculations.
Nevertheless, the calculated spin--averaged masses for both sets of
parameters are close to  each other.  For $P$-wave states the
differences are less than 10 MeV and for $D$-wave states less than 20
MeV (see Table \ref{tab1}).

In many papers excited states of charmonium were studied in the nonrelativistic
approximation which  works quite well for the spectrum. But it was shown
in \cite{8,9} that the relativistic corrections to matrix elements like
$<r^{-3}>$ and $<r^{-3}\ln mr>$, which determine fine structure splittings,
are rather large, about 30$\div$ 40\%. That is why in this paper only
 relativistic calculations of the fine structure  splittings
of the excited states of charmonium will be considered.

\section{Fine structure parameters of the $P$ levels}

Although the spin--averaged masses in our calculations are very close to
 those
in GI's paper \cite{1}, we expect that the spin--orbit and tensor splittings of
the $P$-wave states will be  larger in our case. There are two reasons
for that.  First, we take into account the corrections to second order in
 $\alpha_s$.  Secondly, our calculations of different
matrix elements have shown that for the excited  $P$ states the matrix
element $<r^{-3}>$ which determines the  splittings, is not decreasing.
For the set of parameters (3) it was found
that in the relativistic case $<r^{-3}>_{1P}=0.142, \;
<r^{-3}>_{2P}=0.157, \; <r^{-3}>_{3P}=0.167$, i.e. $<r^{-3}>_{nP}$ are
even increasing for $2P$ and $3P$ states. This result is specific to the
Salpeter equation, whereas in the nonrelativistic case the matrix elements
 $<r^{-3}>$  for the excited states are  decreasing, e.g.
$<r^{-3}>_{1P}=0.101,<r^{-3}>_{2P}=0.093, <r^{-3}>_{3P}=0.089$.
The accuracy of our calculations was  checked to be ($1\div
2)10^{-4}$.

The fine structure parameters are defined as  matrix elements
of the spin-orbit and tensor interactions,
\begin{equation}
a=<\tilde V_{LS}(r)>, \quad c=<\tilde V_T(r)>,
 \label{eq05}
\end{equation}
where
the scalar functions $\tilde V_{LS}(r)$ and $\tilde V_{T}(r)$ are
introduced as
\[
\hat V_{LS}(r) =\tilde V_{LS}(r)\vec L \cdot \vec S,
\]
\begin{equation}
\hat V_{T}(r) =\tilde V_{T}(r)\hat S_{12}, \quad
 \hat S_{12}= 3(\vec s \cdot \vec n) (\vec s_2 \cdot \vec n)
 -\vec s_1 \cdot \vec s_2 , \quad \vec n=\frac{\vec r}{r} .
 \label{eq06}
\end{equation}
Here the spin--orbit parameter $a$ is defined in the same manner as in other
papers, whereas the definition of the tensor parameter $c$ differs from
\cite{1} where the tensor  parameter $T=\frac{1}{2} c$ and from \cite{2} where
the parameter $b$ is used, related to $c$ by $b=4c$.

In our  calculations we suggest that the $P-$wave hyperfine splitting is small 
as it occurs for the $h_c\; (1P)$ meson
for which the hyperfine shift relative to the center of
gravity of the $^3P_J$ multiplet, $\overline M(1^3P_J)$, is less than 1 MeV. 
When the hyperfine
splitting is neglected the mass of the states with spin $S=0$ coincides with
the center of gravity of the $^3L_J$ multiplet denoted by $\overline M_L$. Their
values, taken from Table \ref{tab1},  are 
\[
M(1^1P_1) =3528 \,{\rm MeV}; M(2\,^1P_1) =3962 \,{\rm MeV}
\]
\begin{equation}
M(1^1D_2) =3822 \,{\rm MeV}; M(2\,^1D_2) =4194 \,{\rm MeV},
 \label{eq07}
\end{equation}
which are about 20 MeV lower than in \cite{1} for the $D$ and some $S$
states.

For the states with spin $S=1$ and orbital angular momentum $L\neq 0$ the
mass of the state can be represented  as
\begin{equation}
M(^3L_J)= \overline M_L+a<\vec L  \cdot \vec S>+c<\hat S_{12}>
 \label{eq08}
\end{equation}
with the operator $\hat S_{12}$ defined as in (6).
 For $P$-wave states it
gives
\begin{equation}
M(^3P_2)=\overline M_1+a-\frac{1}{10} c, \quad
M(^3P_1)=\overline M_1-a+ \frac{1}{2} c, \quad
M(^3P_0)=\overline M_1-2a- c.
 \label{eq09}
\end{equation}

In Ref.~\cite{1} only the terms of first  order in $\alpha_s$ were taken
into account and for $a$ and $c$ the following values were obtained,
\[
a(1P)=28 \,{\rm MeV}, \quad c(1P)=26 \,{\rm MeV},
 \label{eq10}
\]
\begin{equation}
a(2P)\approx 17 \,{\rm MeV}, \quad c(2P)\approx 8 \,{\rm MeV},
 \label{eq11}
\end{equation}
which  are about 20\% and 30\% resp. smaller
for  the $1P$ state compared to the  current experimental values \cite{8}:
\begin{equation}
a_{\rm exp}(1P) =34.56 \pm 0.19 \,{\rm MeV} \quad
c_{\rm exp}(1P) =39.12 \pm 0.62 \,{\rm MeV}.
 \label{eq12}
\end{equation}
In our approach the terms of second order   in $\alpha_s$ will be
taken into account and the total value of $a$ and $c$ can be represented as
\begin{equation}
a_{{\rm tot}}=a^{(1)}_P+a_P^{(2)}+a_{\rm NP}, \quad
c_{{\rm tot}}=c^{(1)}_P+c_P^{(2)}+c_{\rm NP},
 \label{eq13}
\end{equation}
where the nonperturbative contribution
to the spin--orbit splitting coming from the linear confining potential is
\begin{equation}
a_{\rm NP}=-\frac{\sigma}{2m^2} <r^{-1}>.
 \label{eq14}
\end{equation}
In the tensor splitting, Eq.~(\ref{eq13}), the small nonperturbative term $c_{\rm NP}$
will be neglected, (see the discussion in \cite{9}). 
Godfrey and Isgur took into account non-perturbative spin effects. In order
to do so, they needed to smear their potentials at short range, which makes it
necessary to intoduce additional unknown parameters.
%To consider perturbative contributions
% some smearing of short range  potentials  at small
%distances  was done in \cite{1} which brings additional unknown
%parameters.  
Here we consider spin--effects as a perturbation using
explicit analytical  expressions for the spin--orbit and tensor
potentials in coordinate space in the $\overline {MS}$ renormalization
scheme from \cite{15}.  The terms to first order in $\alpha_s$ are
$a^{(1)}_P$ and   $c^{(1)}_P$  given by
\begin{equation}
a^{(1)}_P=\frac{2\alpha_s(\mu)}{m^2} <r^{-3}>, \quad
c^{(1)}_P=\frac{4}{3}\frac{\alpha_s(\mu)}{m^2}
 \label{eq15}
\end{equation}
and the second order perturbative corrections are
\[
 a^{(2)}_P=\frac{2\alpha_s^2(\mu)}{\pi m^2} \left\{4.5 \ln \frac{\mu}{m}
<r^{-3}> + 2.5<r^{-3}\ln(mr)>+1.582<r^{-3}> \right\}
\]
\begin{equation}
c^{(2)}_P=\frac{4\alpha_s^2(\mu)}{3\pi m^2}\{4.5 \ln \frac{\mu}{m}
<r^{-3}> + 1.5<r^{-3}\ln(mr)>+3.449<r^{-3}>\}
 \label{eq17}
\end{equation}
The second order expressions are given here for a number of flavors
$n_f=3$. (We checked that for $n_f=4$ the values of
$a$ and $c$  do not practically change--the differences are less than
0.5 MeV--therefore only  the case $n_f=3$ will be presented here.)

With the solutions of the Salpeter equation (1) all matrix elements
determining Eqs~(\ref{eq14}-\ref{eq17}) can
be calculated   and the only uncertainty comes from the
choice of the strong coupling constant $\alpha_s(\mu)$ and the value
of the renormalization scale $\mu$. In \cite{9} it was  found that for
the $1P$ state in charmonium 
\begin{equation}
a_s(\mu)=0.365 \quad (\mu=0.92\, {\rm GeV})
 \label{eq18}
\end{equation}
gives an accurate description of the spin splittings.
For the excited $2P$ and $3P$ states where experimental data are absent,
we shall use the same value (\ref{eq18}) for $\alpha_s(\mu)$. The main
argument in favour of this choice can be taken from the fine  structure
analysis in bottomonium where  the values of $\alpha_s(\mu)$ for the $1P$
and $2P$ states differ by about  10\% only \cite{8}.

With $\alpha_s(\mu) = 0.365$ the authors of \cite{14} obtained for the 
$1P$ state
\[
a^{(1)}_P(1P) = 47.6 \,{\rm MeV}, \quad
a^{(2)}_P(1P) = 3.6 \,{\rm MeV}, \quad
a_{\rm NP}(1P) =-16.6 \,{\rm MeV},
\]
\begin{equation}
c^{(1)}_P(1P) = 31.7 \,{\rm MeV}, \quad
c^{(2)}_P(1P) = 7.4 \,{\rm MeV},
 \label{eq19}
\end{equation}
so that $a_{{\rm tot}}(1P)$ and $c_{{\rm tot}}(1P)$ just coincide with their
experimental values (\ref{eq12}).

For the excited $2P$ state the spin--orbit and tensor parameters are
as large as those of the $1P$ state, because the matrix element
$<r^{-3}>_{2P} $ is  even  $\approx 10$\% larger than $<r^{-3}>_{1P}$.
Here we face the difference between the relativistic
and nonrelativistic approaches. For the latter, matrix elements like
$<r^{-3}>_{nP}$ are decreasing  with growing $n=n_r+1$. Our
calculations give for the $2P$ state that
\[ a^{(1)}_P(2P) = 52.5 \,{\rm MeV}, \quad
a^{(2)}_P(2P) = -0.4 \,{\rm MeV}, \quad
a_{\rm NP}(2P) =-13.4 \,{\rm MeV},
\]
\begin{equation}
c^{(1)}_P(2P) = 35.0 \,{\rm MeV}, \quad
c^{(2)}_P(2P) = 6.5 \,{\rm MeV}
 \label{eq20}
\end{equation}
so that
\begin{equation}
a_{{\rm tot}}(2P)=38.7 \,{\rm MeV}, \quad c_{{\rm tot}}(2P)=41.5 \,{\rm MeV}
 \label{eq21}
\end{equation}
are even slightly larger than  the  corresponding values for the $1P$
state.

Comparing values obtained for $a(2P)$ and $c(2P)$ with those in (11)
from GI's paper  one
can see that $a$ and $c$ in our calculations are larger by a factor of 
2 or 5 resp.  than the values  from GI's paper (see also Table~\ref{tab2}). 
This discrepancy is partly connected to our taking into account of the second 
order radiative corrections which are, however, not large. 
But even with only  first order perturbative terms included,
our  values of $a(2P)$ and $c(2P)$ are much  larger than GI's values (11).

Taking $a$ and $c$ from Eq.~(\ref{eq21}) and $\overline{M}_1 (2P) =3962$ MeV
(from Table~\ref{tab1}) one obtains the following masses for the $2\,^3P_J$ states,
\begin{equation}
M(2\,^3P_0)=3843 \,{\rm MeV} , \quad M(2\,^3P_1)=3944 \,{\rm MeV} ,
 \quad M(2\,^3P_2)=3997 \,{\rm MeV} .
 \label{eq22}
\end{equation}
 Our predicted mass for the $2\,^3P_0$ state, $ M(2\,^3P_0)=3.84$ GeV
appeared to be $\approx 80$ MeV lower than the one in GI's paper, 
whereas for the other two states, $2\,^3P_1$ and $2\,^3P_2$, the predicted masses
 only slightly differ from the GI values (see Table~\ref{tab3}) due to a
 cancelation of terms with different signs.

 It is important  that in our calculation the $2\,^3P_0$ level
lies below the $D\bar D^*$ threshold ($M_{\rm th}(D\bar D^*)\approx 3.87$ GeV) 
but higher than $M_{\rm th}(D\bar D)=3.73$ GeV. This fact can
affect the decay rates of  the $2\,^3 P_0$ state.

For the $3P$ state, using  again $\alpha _s(\mu)=0.365$ and $\mu=0.92$ GeV as 
for the  $1P$ state, one obtains using Eqs.~(\ref{eq14}-\ref{eq17})
\[
a^{(1)}_P(3P) = 56.7 \,{\rm MeV}, \quad
a^{(2)}_P(3P) = -3.6 \,{\rm MeV}, \quad
a_{\rm NP}(3P) =-11.8 \,{\rm MeV},
\]
\begin{equation}
c^{(1)}_P(3P) = 37.8 \,{\rm MeV}, \quad
c^{(2)}_P(3P) = 6.2 \,{\rm MeV}, \;\; (c_{\rm NP}=0),
 \label{eq23}
\end{equation}
 so that
\begin{equation}
a_{{\rm tot}}(3P)=42.3 \,{\rm MeV}, \quad c_{{\rm tot}}(3P)=44.0 \,{\rm MeV}.
 \label{eq24}
\end{equation}
 With these values for $a$ and $c$ and the spin-averaged mass $\overline
 {M}_1(3P)=4320$ MeV it follows that
 \begin{equation}
 {M}(3\,^3P_1)=4300 \,{\rm MeV}, \quad
 {M}(3\,^3P_2)=4358 \,{\rm MeV}, \quad
 {M}(3\,^3P_0)=4192 \,{\rm MeV}.
  \label{eq25}
\end{equation}
 The level $3\,^3P_0$ lies 128 MeV lower than the center of gravity
  of the $3\,^3P_J$ multiplet. It is of interest to notice that the
  difference
 ${M}(n\,^3P_2)-
 {M}(n\,^3P_0)\equiv \Delta (nP)= 3a+0.9 c$
  is large
  in all cases and slightly increasing for the  excited states,
\begin{equation}
  \Delta(1P) =138.9 \,{\rm MeV}, \quad \Delta (2P)=143 \,{\rm MeV},
 \quad \Delta (3P)= 166 \,{\rm MeV} .
  \label{eq26}
\end{equation}

We have checked the sensitivity of the predicted values for $a$ and $c$
to the choice of the renormalization scale $\mu$ and
$\alpha_s(\mu)$. To this end the commonly used value of $\mu=m$ and
$\alpha_s(\mu=m=1.48\, {\rm GeV})=0.29$ was considered. Then for the $2P$ state
$a_{{\rm tot}}(2P)=36.5 $\,{\rm MeV} and $c_{{\rm tot}}(2P)=37.5 $\,{\rm MeV} 
and for the $3P$ state $a_{{\rm tot}}(3P) =40.5 $\,{\rm MeV} and 
$c_{{\rm tot}} (3P) =39.8 $\,{\rm MeV} were obtained,
which  are very close to the values (20), (21), (23), and (24) with
$\mu_0=0.92$ GeV and $\alpha_s(\mu_0) =0.365$,
found in \cite{8} from the fit to the $1P$ fine structure splittings.

\section{Fine structure splittings of the $D$ levels}

For the $D$
state the expressions of the splitted masses, $M(n\,^3D_J)$, through the
parameters $a$ and $c$ can be found in \cite{2},
\begin{equation}
 M(\,^3D_1)=\overline M_2-3a-\frac{1}{2}c, \quad
 M(^3D_2)=\overline M_2-a+\frac{1}{2}c,\quad
 M(^3D_3)=\overline M_2+2a-\frac{1}{7}c.
 \label{eq27}
\end{equation}
For the spin-averaged masses $\overline{M}(nD)$ our calculation
with the parameters (3) give
\begin{equation}
 \overline{M}_2(1D)=3822 \,{\rm MeV}, \quad \overline{M}_2(2D)=4194 \,{\rm MeV}.
 \label{eq28}
\end{equation}

The fine structure parameters $a$ and $c$ for the $D$-wave levels are
given in Table~\ref{tab2} together with their values from GI's paper. As seen
from Table~\ref{tab2}, $a$ and $c$ in both cases practically coincide. Still
our predicted masses for the $1^3D_J$  and $2\,^3D_J$ states appear to be
about 20 MeV lower than in \cite{1} because of the smaller value of the
spin--averaged masses.

\section{Conclusion}

In our analysis it was found that

   (i) in the relativistic case the fine  structure splittings of the 
charmonium $P$-states with spin $S=1$ are much larger than in
the nonrelativistic case;

 (ii)  for the excited $2P$ states the parameters of the fine structure
are even slightly larger than for the $1^3P_J$ ground state;

(iii)  the mass of the $n\,^3P_0$ states $(n=1,2,3)$ appears to be about
130 MeV smaller than the  center of gravity of the $n\,^3P_J$ multiplet.
This  fact can be  important for the explanation of the decays of
the excited states of charmonium.

The value we predict for $M(2\,^3P_0)=3.84$ GeV is about 80 MeV lower
than in GI's calculations \cite{1}. This state lies below  the $D\bar D^*$
threshold and  is only about 100 MeV higher than the $D\bar D$
threshold. The point of view exists that this state could be very
broad because it lies above the $D\bar D$ threshold and therefore
should have a large hadronic width. On the  other  hand this state
lies relatively close to the $D\bar D$ threshold, so the width could be 
suppressed by phase space limitations.
%and the suppression of phase space could take place. 
Therefore this state could play a
role  in the production of $\psi(2S)$ charmonium mesons as it was
discussed in \cite{5,6}.

   \newpage

\begin{table}
 \begin{center}
 \caption{ The spin-average masses $M(nL)$ (in  MeV) in charmonium
           for two sets of parameters.  \label{tab1}}

\vspace{1ex}

   \begin{tabular}{|l|l|l|l|}
   \hline
   &Godfrey and Isgur& present paper&\\
  $ M(nL)$ &$ m=1.628 {\rm GeV}$& $m=1.48\, {\rm GeV}$& \\
           &$\sigma= 0.18 {\rm GeV}^2$&$\sigma= 0.18 {\rm GeV}^2$& Experiment\\
    &$\tilde \alpha_{crit}=0.6$ (running $\tilde \alpha(r)$)&
    $\tilde \alpha=0.42$,&\\
    &$C_0=-253 $ MeV & $C_0=-140$ MeV  & \\
    \hline
    $\psi (1S)$& 3067.5& 3067.6 & 3067.6$\pm$0.6\\ \hline
    $\psi (2S)$& 3665  & 3659$^{a)}$ & 3663$\pm$1.3\\ \hline
    $\psi (3S)$& 4090  & 4077$^{b)}$ & 4040$\pm$10 \\ \hline
    $\psi (4S)$& 4450  & 4425        & 4415$\pm$6  \\ \hline
    $\psi (5S)$&   & 4732        &             \\ \hline
    $\chi_c (1P)$& 3520  & 3528        & 3525.5$\pm$0.4  \\ \hline
    $\chi_c (2P)$& 3960  & 3962        &                 \\ \hline
    $\chi_c (3P)$&       & 4320        &                 \\ \hline
    $M(1D)$& 3840  & 3822        & 3768.9$\pm$2.5  \\ \hline
    $M(2D)$& 4210  & 4194        & 4159  $\pm$20   \\ \hline
    $M(3D)$& 4520  & 4519.5      &              \\  \hline
    \end{tabular}
    \end{center}

    $^{a)}$ Mixing of $2S$ and $1D$-wave states is not taken into
    account

    $^{b)}$ Mixing of $3S$ and $2D$-wave states is not taken into
    account
\end{table}

\begin{table}
 \begin{center}
 \caption{Spin-orbit and tensor splittings $a$ and $c$ (in MeV) for
   $P$ and $D$ levels.\label{tab2}} 

\vspace{1ex}

   \begin{tabular}{|l|l|l|l|}
   \hline
   &Godfrey and Isgur& present paper&\\
    & paper$^{a)}$ & $\alpha_s(\mu)= 0.365$&
     Experiment\\
    \hline
    $a(1P)$& 28    & 34.56  &34.56  $\pm$0.19\\ \hline
    $c    (2P)$& 26    & 39.12       & 39.12$\pm$0.62\\ \hline
    $a (2P)$& 17  & 38.7 &  \\ \hline
    $c (2P)$& 8  & 41.5        &   \\ \hline
    $a (3P)$&   & 42.3        &   \\ \hline
    $c (3P)$&   & 44.0        &   \\ \hline
    $a (1D)$&$\approx$ 5   & 3.64        &   \\ \hline
    $c (1D)$&$\approx$ 10   & 10.94        &   \\ \hline
    $a (2D)$&$\approx$ 5   & 5.43          &   \\ \hline
    $c (2D)$&$\approx$ 10   & 11.37          & \\  \hline
    \end{tabular}
    \end{center}

    $^{a)}$ the values of $a,c$ for $2P$ and $D$ states are
    extracted from the masses $M(^3D_J)$ and $M(2^3P_J)$ given in
    Ref.~\cite{1}.
\end{table}

\begin{table}
 \begin{center}
 \caption{\label{tab3} Masses of the $n^3P_J$, and $n^3D_J$ states (in MeV) in 
            charmonium.}

\vspace{1ex}

   \begin{tabular}{|l|l|l|l|}
   \hline
   &Godfrey and Isgur& present paper&\\
    &paper&$\alpha_s(\mu)= 0.365$&
     Experiment\\
    \hline
    $2^3P_0$& 3920    & 3843  &  \\
    $2^3P_1$& 3950    & 3944  &  \\
    $2^3P_2$& 3980    & 3996  &  \\ \hline
    $3^3P_0$&     & 4192  &  \\
    $3^3P_1$&     & 4300  &  \\
    $3^3P_2$&     & 4358  &  \\ \hline
    $1^3D_1$& 3820    & 3800$^{a)}$    & 3768.9$\pm$2.5\\
    $1^3D_2$& 3840    & 3823    & \\
    $1^3D_3$& 3850    & 3827    & \\ \hline
    $2^3D_1$& 4190   & 4167$^{b)}$    & 4159$\pm$20\\
    $2^3D_2$& 4210    & 4195   & \\
    $2^3D_3$& 4220    & 4204    & \\ \hline
    \end{tabular}
    \end{center}

    $^{a)}$ see footnote $^{a)}$ to Table \ref{tab1};
    $^{b)}$ see footnote $^{b)}$ to Table \ref{tab1}.
\end{table}

\end{document}